\documentclass[12pt,a4paper]{article}
\usepackage[truedimen,margin=30mm]{geometry} 
\usepackage{mathrsfs}
\usepackage{amssymb}
\usepackage{amsmath}
\usepackage{ascmac}
\usepackage{amsthm}
\usepackage[dvipdfmx]{graphicx}
\usepackage{natbib}
\usepackage{setspace}
\usepackage{times}

\usepackage{color}
\usepackage{float}

\usepackage{booktabs}
\usepackage{multirow}
\usepackage{threeparttable}
\usepackage{siunitx}

\usepackage{tikz}
\usepackage{titlesec}
\titleformat*{\section}{\large\bfseries}
\titleformat*{\subsection}{\it}

\DeclareMathOperator*{\argmin}{arg\,min}

\usetikzlibrary{shapes.geometric, arrows, positioning, calc, fit, backgrounds}
\definecolor{tree1color}{RGB}{70, 130, 180}
\definecolor{tree2color}{RGB}{205, 92, 92}
\definecolor{sumcolor}{RGB}{60, 179, 113}
\definecolor{leafbg}{RGB}{240, 248, 255}

\title{{\bf Direct Bayesian Additive Regression Trees for Conditional Average Treatment Effects in Regression Discontinuity Designs}\footnote{\today. Corresponding author: Shonosuke Sugasawa (Email: sugasawa@econ.keio.ac.jp)}
}

\date{}

\begin{document}

\maketitle
\doublespacing

\vspace{-1.5cm}
\begin{center}
{\large Daisuke Kondo$^1$ and Shonosuke Sugasawa$^1$}

\medskip
%\today

\medskip
\noindent
$^1$Faculty of Economics, Keio University\\
\end{center}

\medskip
\begin{center}
{\bf Abstract}  
\end{center}

Regression discontinuity designs (RDD) are widely used for causal inference. In many empirical applications, treatment effects vary substantially with covariates, and ignoring such heterogeneity can lead to misleading conclusions, which motivates flexible modeling of heterogeneous treatment effects in RDD. To this end, we propose a Bayesian nonparametric approach to estimating heterogeneous treatment effects based on Bayesian Additive Regression Trees (BART). The key feature of our method lies in adopting a general Bayesian framework using a pseudo-model defined through a loss function for fitting local linear models around the cutoff, which gives direct modeling of heterogeneous treatment effects by BART. Optimal selection of the bandwidth parameter for the local model is implemented using the Hyv\"arinen score. Through numerical experiments, we demonstrate that the proposed approach flexibly captures complicated structures of heterogeneous treatment effects as a function of covariates.

\bigskip\noindent
{\bf Key words}: general Bayes; Bayesian additive regression trees; local estimation; uncertainty quantification

\newpage
%----------------------------------------%
%           Introduction                 %
%----------------------------------------%
\section{Introduction}

Regression discontinuity design (RDD) is a representative method for causal inference under intervention determined by a continuous (running) variable and is widely adopted in a variety of areas such as economics, marketing, epidemiology, and clinical research. 
RDD was introduced by \citet{ThistlethwaiteCampbell1960RDD}, and \citet{hahn2001identification} formalized identification conditions and proposed local linear estimators. 
Subsequent methodological developments have focused on improving estimation and inference for the average treatment effect at the cutoff, including comprehensive surveys of the literature \citep{ImbensLemieux2008RDDGuide, LeeLemieux2010RDDSurvey}, optimal bandwidth selection \citep{ImbensKalyanaraman2012IK}, robust confidence intervals \citep{CalonicoCattaneoTitiunik2014RDDCI}, and Bayesian nonparametric approaches \citep{ChibGreenbergSimoni2023NonparametricRD}.
Despite these methodological advances, the majority of the RDD literature has focused on estimating average treatment effects (ATE) on the treated at the cutoff.
However, in many empirical applications, treatment effects are likely to vary substantially across units with different pretreatment covariates \citep[e.g.][]{becker2013absorptive,card2008impact}. 
Ignoring heterogeneity can obscure important distributional features of the treatment effect and lead to misleading conclusions.

To estimate conditional average treatment effect (CATE) in RDD, several approaches have been proposed in the literature. 
\citet{becker2013absorptive} and \citet{calonico2025treatment} employ local polynomial estimation and estimate CATE under parametric specifications. 
\citet{sugasawa2023hierarchical} and \citet{tao2025bayesian} propose subgroup-wise ATE estimation, and \citet{sawada2024local} addresses treatment effect heterogeneity across different regions of the running variable. 
These methods, however, do not provide CATE as a smooth function of covariates. 
\citet{reguly2021discovering} proposes a tree-based method in which the data are recursively partitioned using covariates, and standard RDD estimation is conducted within each leaf. 
While this provides a nonparametric approach to CATE, a single tree may lack sufficient flexibility, especially in the presence of many covariates, and coherent uncertainty quantification of the CATE estimator may be challenging. 
As the most closely related work, \citet{alcantara2025learning} propose the use of Bayesian additive regression trees (BART) to estimate the conditional expectation of the outcome as a function of covariates and the running variable while accounting for the threshold.
Although it enables flexible estimation of CATE, it models the entire conditional outcome surface and derives the treatment effect indirectly as a contrast. 
As a result, nuisance components unrelated to the treatment effect must also be modeled through BART and response models may not adequately control regularization on CATE itself.
It has been recognized in the Bayesian causal inference literature that modeling the entire response surface without directly regularizing heterogeneous treatment effects may lead to biased or unstable estimation, and that separate prior structures targeting CATE can improve performance \citep[e.g.,][]{hahn2020bayesian}.

This paper introduces an alternative Bayesian approach using BART for estimation and inference on CATE in RDD.  
The key feature of our method lies in adopting a general Bayesian framework using a pseudo-model defined through a loss function for fitting local linear models around the cutoff.
This formulation enables direct modeling of CATE through BART, which we call ``Direct-BART", and fully exploits the flexibility of BART to capture unknown and potentially complex forms of heterogeneity.
The posterior computation of Direct-BART can be performed by a simple Gibbs sampling where the backfitting algorithm for BART can be readily incorporated and the sampling step for the local linear regression coefficients is straightforward.    
While the proposed method includes a tuning parameter which determines the local weights around the threshold, we propose a data-driven approach to bandwidth selection based on the Hyv\"{a}rinen score \citep{HyvarinenDayan2005ScoreMatching,YonekuraSugasawa2023GeneralBayesRobust}, which can be efficiently evaluated using posterior samples.
We demonstrate the Direct-BART through simulation experiments and show that it provides superior performance to the existing approaches.

The remainder of the paper is organized as follows.
Section~\ref{sec:Method} introduces the proposed approach, including the posterior computation algorithm and a data-driven bandwidth selection based on the Hyv\"{a}rinen score.
Section~\ref{sec:Simulation} presents simulation studies to assess the performance of the proposed approach in comparison with several methods.
Concluding remarks are given in Section~\ref{sec:Concluding Remarks}.

%----------------------------------------%
%              Method                    %
%----------------------------------------%
\section{BART for Direct Inference on CATE in RDD}
\label{sec:Method}

\subsection{Pseudo-model for direct inference on CATE}
We consider the standard setup of a sharp RDD.
Let $Y_i$ denote the outcome variable and $X_i$ the running variable
for unit $i=1,\ldots,n$.
The treatment indicator is defined as $W_i = I(X_i \ge c)$, where $c$ is a known cutoff value.
We additionally observe a $d$-dimensional vector of covariates $Z_i$, which is used to model treatment effect heterogeneity.
The CATE at the cutoff, conditional on $Z_i=z$, is defined as
$$
\tau(z)=\lim_{x \downarrow c} \mathbb{E}[Y_i \mid X_i = x, Z_i = z]
-\lim_{x \uparrow c} \mathbb{E}[Y_i \mid X_i = x, Z_i = z].
$$

A standard approach of RDD assuming homogeneity of the treatment effect is to fit local linear (polynomial) regression as a function of the running variable.
While the local regression is typically fitted to treatment and control regions separately, one can fit the weighted local linear regression to both regions simultaneously by minimizing the following loss function:
\begin{align*}
\frac{1}{2} \sum_{i=1}^n
K\!\left(\frac{|X_i-c|}{h}\right)
\left(Y_i - \tau W_i - X_i(c)^\top \beta \right)^2,
\end{align*}
where $X_i(c)=\bigl( 1,(X_i-c)_-,(X_i-c)_+,\ldots,(X_i-c)_-^q,(X_i-c)_+^q\bigr)^\top$ is a vector for polynomial regression, $K(\cdot)$ is a kernel function, and $h$ is a bandwidth parameter.
\cite{sugasawa2023hierarchical} adopted the above loss function to estimate group-wise ATE. 

Since we allow for heterogeneous treatment effects $\tau(Z_i)$, the local regression framework must also accommodate covariate-dependent coefficients. In particular, under heterogeneity, the local regression coefficient $\beta$ naturally becomes a function of $Z_i$, which we denote by $\beta(Z_i)$.
To flexibly capture this dependence while maintaining analytical and computational tractability, we assume that the varying coefficient can be approximated by a linear function of $Z_i$, namely, $\beta(Z_i)=B\tilde{Z}_i$, where $\tilde{Z}_i=(1,Z_{i1},\ldots,Z_{id})^{\top}$ and $B\in\mathbb{R}^{(2q+1)\times(d+1)}$ is a coefficient matrix.
This specification can be viewed as a natural extension of the conventional local linear RDD framework, in which $\beta$ is assumed to be constant across units. 
Here, we retain the linear structure underlying standard local regression methods while allowing the regression coefficients to vary systematically with observed covariates.
Hence, we consider the following loss function:
\begin{align}
L(\tau, B; h) = \frac{1}{2} \sum_{i=1}^n
K\!\left(\frac{|X_i-c|}{h}\right)
\left(Y_i - \tau(Z_i)W_i - X_i(c)^\top B \tilde{Z}_i \right)^2,
\label{eq:loss}
\end{align}
where $\tilde{Z}_i = (1, Z_{i1}, \ldots, Z_{id})^{\top}$. 
For the kernel function, we employ the uniform kernel $K(x)=I(|x|\le 1)$ throughout the paper.

To implement the loss function (\ref{eq:loss}) in a Bayesian framework, we consider a pseudo statistical model for
$Y=(Y_1,\ldots,Y_n)$, defined as $p_h(Y \mid \tau, B,\omega)
\propto \exp\left\{-\omega L(\tau, B; h)\right\}$, where $\omega>0$ is a universal scaling parameter.
The use of a pseudo-model to define posterior distributions is known as a general Bayesian method \citep[e.g.][]{Bissiri2016GeneralFramework}.
The pseudo-model can be equivalently expressed as a power likelihood of a gaussian model:
\begin{equation}
p_h(Y_i \mid \tau,B,\omega)
\propto \phi(Y_i;\tau(Z_i)W_i + X_i(c)^\top B\tilde{Z}_i, \omega^{-1})^{k_i},
\label{eq:powerlikelihood}
\end{equation}
independently for $i=1,\ldots,n$, where $k_i = K\left({|X_i-c|}/{h}\right)$.
This model enables us to assign BART prior to $\tau(Z)$ directly, which will be discussed in the subsequent section.

For the local regression coefficient matrix and the scaling parameter, we employ priors $B \sim \mathcal{MN}_{2q+1,\,d+1}\left(M_0,V_0,U_0\right)$ and $\omega \sim \mathrm{Ga}(\nu_0,\eta_0)$, where $M_0, U_0, V_0$, $\nu_0$, and $\eta_0$ are fixed hyperparameters. Here, $\mathcal{MN}_{2q+1,\,d+1}(M_0,V_0,U_0)$ denotes a matrix normal distribution
with mean matrix $M_0$ and row and column covariance matrices $V_0$ and $U_0$, respectively, and $\mathrm{Ga}(\nu_0,\eta_0)$ denotes a Gamma distribution
with shape parameter $\nu_0$ and rate parameter $\eta_0$. 
As a default choice, we set $M_0 = \mathbf{0}_{2q+1,d+1},V_0 = 100\mathbf{I}_{2q+1},U_0 = \mathbf{I}_{d+1}, \nu_0 = 1, \eta_0 = 1$, which will be used in our numerical studies.

\subsection{BART prior for CATE}

Bayesian Additive Regression Trees (BART) is a nonparametric prior on the regression function introduced by \citet{Chipman2010BART}, which can capture complex and nonlinear functions of covariates. 
We here directly introduce the BART prior for the CATE function $\tau(Z)$. 
Hence, the CATE function is modeled as $\tau(Z_i)=\sum_{j = 1}^m g(Z_i\mid T_j, M_j)$, where $m$ is the number of trees, $T_j$ denotes the structure of the $j$-th regression tree including the splitting rules, and $M_j = (\mu_{j1}, \mu_{j2},\ldots, \mu_{jb_j})$ denotes the corresponding set of leaf parameters. 
Each tree $T_j$ partitions the covariate space into a collection of disjoint regions $\{G_{j1}, G_{j2}, \ldots, G_{jb_j}\}$ through its splitting rules and each leaf parameter is assigned for each region. 
If a covariate vector $z$ falls into a covariate region $G_{jk}$, the function $g(z\mid T_j, M_j)$ returns the associated leaf parameter $\mu_{jk}$, that is, $g(z\mid T_j, M_j) = \mu_{jk}$ if $z \in G_{jk}$ for $k = 1,2,\ldots, b_j$. 
By aggregating all trees, BART partitions the covariate space into fine disjoint regions, given by intersections of tree-specific regions such as $\bigcap_{j=1}^m G_{j k_j}$, which allows BART to flexibly estimate the CATE function.

The parameters in BART are $(T_j, M_j)_{j=1}^m$, and we assign independent priors to the tree components $(T_j, M_j)$. 
Moreover, the leaf parameters within each tree are assumed to be independent a priori. Accordingly, we specify priors as $p(T_j, M_j)=p(T_j)\prod_{k=1}^{b_j} p(\mu_{jk}\mid T_j)$. 
The tree prior $p(T_j)$ consists of three components: (i) the probability that a node splits, (ii) the probability to choose which variable to split, and (iii) the probability to choose the threshold value after choosing a variable to split. For component (i), to encourage shallow tree structures, the prior probability that a node at depth $d \ (=0,1,2,\ldots)$ splits is given by $\alpha(1+d)^{-\beta}$ for $\alpha\in(0,1)$ and $\beta \in [0,\infty)$, where $\alpha = 0.95, \beta = 2$ is the default choice, following the standard BART specification \citep[e.g.][]{Chipman2010BART,linero2018bayesian}. 
For components (ii) and (iii), the splitting variable and the threshold are selected uniformly from their respective candidate sample sets. 
The prior on a leaf parameter is a normal distribution, specifically, $p(\mu_{jk}\mid T_j)\sim N(\mu_{\mu}, \sigma^2_{\mu})$ independently. 
The prior on error variance $p(\sigma^2)$ is an inverse-Gamma distribution.
In contrast, we explicitly specify the prior mean and standard deviation of the leaf parameters, denoted by $\mu_\mu$ and $\sigma_\mu$, respectively.
To set $\mu_\mu$ and $\sigma_\mu$, we follow a strategy similar to that of \citet{Chipman2010BART}.

Given the definition of the treatment effect, it is reasonable to assume that the treatment effect lies within an interval $(\tau_{\min}, \tau_{\max})$, where
$$
\tau_{\max}
= \max_{i:x_i\in (c, c+\delta)} Y_i
  - \min_{i:x_i\in (c-\delta, c)} Y_i,
\ \ \ 
\tau_{\min}
= \min_{i:x_i\in (c, c+\delta)} Y_i
  - \max_{i:x_i\in (c-\delta, c)} Y_i,
$$
for some $\delta > 0$.
Here, $\tau_{\max}$ represents an empirical upper bound on the treatment effect, defined as the difference between the maximum outcome among treated units and the minimum outcome among control units in a $\delta$-neighborhood of the cutoff.
Similarly, $\tau_{\min}$ represents an empirical lower bound on the treatment effect, defined as the difference between the minimum outcome among treated units and the maximum outcome among control units in a $\delta$-neighborhood of the cutoff.

Since the BART prior with $m$ trees implies that the treatment effect follows
a normal distribution $\mathcal N(m\mu_{\mu},\, m\sigma^2_{\mu})$,
we set $\mu_\mu$ and $\sigma_\mu$ so as to assign high prior probability to the interval $(\tau_{\min}, \tau_{\max})$. This is achieved by setting
$m\mu_{\mu} + k\sqrt{m}\sigma_{\mu} = \tau_{\max}$ and
$m\mu_{\mu} - k\sqrt{m}\sigma_{\mu} = \tau_{\min}$
for a suitable constant $k>0$.
Accordingly, we set
$$
\mu_{\mu} = \frac{\tau_{\max} + \tau_{\min}}{2m}, 
\ \ \ \ 
\sigma_{\mu} = \frac{\tau_{\max} - \tau_{\min}}{2k\sqrt{m}}
$$
In our implementation, we set $\delta = 0.1\,\mathrm{sd}(X)$ and $k = 2$, which corresponds to a narrow neighborhood around the cutoff and assigns approximately $95\%$ prior probability mass to the empirical treatment effect interval.

A key feature of our formulation is that the BART prior is placed directly on the CATE function $\tau(Z)$ rather than on the full conditional outcome regression surface. 
This implies that the regularization induced by the prior acts directly on the heterogeneous treatment effect, without being mediated through nuisance components unrelated to the causal estimand. 
By modeling $\tau(Z)$ explicitly, our specification of $(\mu_\mu, \sigma_\mu)$ through $(\tau_{\min}, \tau_{\max})$ enables transparent prior elicitation in terms of plausible magnitudes of treatment effects. 
Moreover, such targeted regularization can be particularly important in high-dimensional settings, where modeling the entire response surface may lead to biased shrinkage for the treatment effect itself, as emphasized in the context of CATE estimation \citep[e.g.,][]{hahn2020bayesian}.

\subsection{Posterior Computation}

Posterior inference of the proposed Direct-BART is conducted via Gibbs sampling, using the backfitting scheme given in the standard BART framework \citep{Chipman2010BART}.
For notational convenience, we write $\tilde{X}_i$ in place of $X_i(c)$ throughout this section.
The joint posterior of $\{B, \omega, (T_j, M_j)_{j=1}^m\}$ is proportional to 
\begin{align*}
p(B)p(\omega)\prod_{j=1}^m p(T_j, M_j)
\prod_{i=1}^n \phi\bigg(Y_i; W_i\sum_{j=1}^m g(Z_i\mid T_j, M_j) + \tilde{X}_i^\top B\tilde{Z}_i, \omega^{-1}\bigg)^{k_i},
\end{align*}
where $p(B)$, $p(\omega)$ and $p(T_j, M_j)$ are prior distributions, described in the previous section. 

Given the CATE function $\tau(Z_i)\equiv \sum_{j=1}^m g(Z_i\mid T_j, M_j)$, namely, given $(T_j, M_j)_{j=1}^m$, the full conditional distributions of $B$ and $\omega$ are familiar forms and their sampling is straightforward. 
Specifically, the full conditional of $B$ is $\mathcal{MN} (M_n, \Sigma_n)$, where
\begin{align*}
M_n &= \Sigma_n\Big[ 
\omega\sum_{i=1}^n k_i \Big\{ Y_i - W_i\sum_{j=1}^mg(Z_i\mid T_j, M_j) \Big\} 
(\tilde{Z}_i \otimes \tilde{X}_i) +\Sigma_0 \operatorname{vec}(M_0) 
\Big]\\
\Sigma_n &= \Big{\{}\omega\sum_{i=1}^n k_i(\tilde{Z}_i \otimes \tilde{X}_i)(\tilde{Z}_i \otimes \tilde{X}_i)^{\top} + \Sigma_0\Big{\}}^{-1},
\end{align*}
and that of $\omega$ is $\mathrm{Ga}(\nu_n, \eta_n)$, where $\nu_n = \nu_0 + \sum_{i=1}^nk_i/2$ and 
\begin{align*}
\eta_n &=\eta_0 + \frac12\sum_{i=1}^nk_i\Big\{ Y_i - W_i\sum_{j=1}^mg(Z_i\mid T_j, M_j) - \tilde{X}_i^{\top}B\tilde{Z}_i \Big\}.
\end{align*}
Regarding parameters $(T_j, M_j)_{j=1}^m$ in BART, they are updated in a sequential way, known as ``Bayesian backfitting".
Specifically, for $j=1,\ldots,m$, we generate $(T_j,M_j)$ from the full conditional $p(T_j, M_j\mid (T_h,M_h)_{h\neq j}, B,\omega, y)$ using the Metropolis–Hastings algorithm, 
where $y=(Y_1,\ldots,Y_n)$ denotes the observed responses for the $n$ units.
To avoid complexities arising from the varying dimension of $M_j$, we first sample $T_j$ from the marginal distribution $p(T_j\mid (T_h,M_h)_{h\neq j},B,\omega,y)$ and then sample $M_j$ from $p(M_j\mid T_j,(T_h,M_h)_{h\neq j}, B,\omega, y)$.
To draw a sample from $p(T_j\mid (T_h,M_h)_{h\neq j}, B,\omega,y)$, we generate a proposal $T_j^{\ast}$ following the strategy of \citet{Chipman2010BART} with the corresponding acceptance probability.
Note that the log-likelihood $\log l_{j,b}$ for a leaf node $b$ in $T_j$ is given by
\begin{align*}
\log l_{j,b}=
&-\frac{n_b}{2}\log(2\pi) + \frac{n_b}{2}\log (\omega) 
-\frac{1}{2}\log(\sigma_\mu^2)
-\frac{1}{2}\log\!\left( \sigma_\mu^{-2}
 + \omega\sum_{i\in\{i;Z_i \in G_{jb}\}}k_i
\right)
\\
& \ \ 
+\frac{1}{2}
\left\{
\frac{
\left(
\sigma_\mu^{-2}\mu_\mu
+ 
\omega\sum_{i\in\{i;Z_i \in G_{jb}\}}k_i R_{ji}
\right)^2
}{
\sigma_\mu^{-2}+
\omega\sum_{i\in\{i;Z_i \in G_{jb}\}}k_i}
-\omega\sum_{i\in\{i;Z_i \in G_{jb}\}} k_i R_{ji}^2
-
\frac{\mu_\mu^2}{\sigma_\mu^2}
\right\},
\end{align*}
and $R_{ji} = Y_i - \sum_{h\neq j} g(Z_i\mid T_h, M_h)- X_i(c)^\top B\tilde{Z}_i$ is a residual and $n_b$ denotes the number of observations assigned to region $G_{jb}$, that is, $n_b = \#\{ i:Z_i\in G_{jb}\}$.
After drawing a sample of $T_j$, we sample leaf parameter $\mu_{jb}$ contained in $M_j$ from $\mathcal{N}(m_n, \sigma^2_n)$, where
\begin{align*}
m_n &= \frac{\sigma^{-2}_{\mu}\mu_{\mu} + \omega\sum_{i\in\{i;Z_i \in G_{jb}\}}k_i R_{ji}}{\sigma^{-2}_{\mu} + \omega\sum_{i\in\{i;Z_i \in G_{jb}\}}k_i }, \qquad
\sigma^2_n = \frac{1}{\sigma^{-2}_{\mu} + \omega\sum_{i\in\{i;Z_i \in G_{jb}\}}k_i}. 
\end{align*}

Based on the posterior samples, we can approximate the posterior distribution of $\tau(z)$ for arbitrary covariate $z$. 
This provides point estimates by computing posterior means as well as uncertainty measures by deriving, for example, credible intervals of $\tau(z)$.

\subsection{Bandwidth Adaptation}

As in local polynomial regression, the choice of the bandwidth parameter $h$ plays a crucial role in RDD. 
In particular, under the proposed model, the bandwidth parameter determines not only the bias--variance tradeoff of the local polynomial approximation, but also the amount of data effectively utilized by the BART prior. 
To select $h$, we consider the leave-one-out pseudo predictive distribution for the out-of-sample observation based on the pseudo-model, defined as $p_h(Y_{i}\mid Y_{-i})=\int p_h(Y_{i}\mid \theta)p_h(\theta\mid Y_{-i})d\theta$, where $Y_{-i}$ denotes the set of observations except for $Y_i$, $\theta \equiv (\tau,B,\omega)$ is a set of parameters, $p_h(Y_i\mid\theta)$ is defined in \eqref{eq:powerlikelihood}, and $p_h(\theta\mid Y_{-i})$ denotes the posterior distribution of $\theta$ given $Y_{-i}$ and bandwidth $h$.
Since the pseudo predictive distribution $p_h(Y_i\mid Y_{-i})$ is not necessarily normalized, we evaluate the predictive distribution by the Hyv\"{a}rinen score \citep{HyvarinenDayan2005ScoreMatching,ShaoJacobDingTarokh2019Hyvarinen} as a function of $h$. 
In particular, following \cite{sugasawa2023hierarchical}, we propose the following local Hyv\"{a}rinen score to focus the evaluation on the neighborhood of the cutoff:
$$
H(h)=\sum_{i\in N_s} \left[2\frac{\partial^2}{\partial Y_i^2}\log p_h(Y_{i}\mid Y_{-i})+\left(\frac{\partial}{\partial Y_i}\log p_h(Y_i\mid Y_{-i})\right)^2\right],
$$
where $N_s$ is the evaluation set defined as the $s$-nearest observations to the threshold.
We set $s$ to $\max(0.02n, 5)$ as a default choice, which will be used in our numerical studies. 
As noted in \cite{YonekuraSugasawa2023GeneralBayesRobust}, the above criterion can be expressed as 
\begin{align}
H(h)
=\sum_{i\in N_s} 
\bigg\{
2 \mathbb{E}\Big[\ell^{(2)}_h(\theta) + \big\{\ell^{(1)}_h(\theta)\big\}^2\Big] 
-\Big(\mathbb{E}\left[\ell^{(1)}_h(\theta)\right]\Big)^2\bigg\},
\label{H-score}
\end{align}
where the expectation is taken with respect to the full sample posterior distribution $p_h(\theta\mid Y)$ and $\ell^{(k)}_h(\theta)\equiv \partial^{k}\log p_h(Y_i\mid \theta)/\partial Y_i^k$ for $k=1,2$.
From (\ref{eq:powerlikelihood}), it follows that 
$$
\ell^{(1)}_h(\theta)= -\omega k_{i}(Y_i-\tau(Z_i) W_i-X_i(c)^\top B \tilde{Z}_i), \ \ \ \ \ 
\ell^{(2)}_h(\theta)= -\omega k_i.
$$
To select the bandwidth parameter $h$ based on the Hyv\"{a}rinen score \eqref{H-score},
we consider a finite set of candidate bandwidth values $\{a_1,\ldots,a_L\}$.
For each candidate $a_\ell$, we fix $h=a_\ell$ and run the MCMC algorithm
for a short batch of iterations to compute the corresponding Hyv\"{a}rinen score $H(a_\ell)$.
We then select the bandwidth by $\hat h = \argmin_{1 \leq \ell \leq L} H(a_\ell)$.

%----------------------------------------%
%             Simulation                 %
%----------------------------------------%
\section{Simulation Study}
\label{sec:Simulation}

We demonstrate the performance of the proposed Direct-BART together with some existing methods through simulation studies based on two scenarios of data-generating processes.

\subsection{Data Generating Processes}

Both scenarios share the following common structural form for outcome: 
$$
Y_i = \mu(X_i, Z_i) + W_i \tau(Z_i) + \varepsilon_i,
\qquad
\varepsilon_i \sim \mathcal N(0,\sigma^2),
$$
where $W_i = I(X_i \geq c)$ denotes the treatment indicator with cutoff value $c=0$. 
The two scenarios differ in the distributions of the running variable $X_i$ and covariates $Z_i$, as well as in the specifications of the baseline function $\mu$ and the treatment effect function $\tau$.

\paragraph{{\bf (Scenario 1)}}  
We set $n=1200$ and the first and second halves of the sample correspond to control and treated units, respectively. 
The running variable $X_i$ is generated from $\mathrm{U}(-1,0)$ for control units and from $\mathrm{U}(0,1)$ for treated units. 
The five-dimensional covariate vector $Z_i=(Z_{i1},\ldots,Z_{i5})^\top$ consists of four continuous variables and one categorical variable, where $
(Z_{i1},\ldots,Z_{i4})^\top \sim
\mathcal N_4 ((\gamma_0+\gamma_1 X_i)\mathbf 1_4,\Sigma)$ and $\mathbf 1_4$ is a four-dimensional vector of ones.
We set $(\gamma_0,\gamma_1)=(-1,0.55)$, yielding $\mathrm{Cor}(X_i,Z_{i1})\approx0.3$.
The $(j,k)$-element of the covariance matrix $\Sigma$ is $\Sigma_{jk}=\{1+|j-k|\}^{-1}$ for $j,k\in\{1,2,3,4\}$, representing moderate correlations. 
The categorical covariate $Z_{i5}\in\{1,2,3\}$ is generated as $\Pr(Z_{i5}=k)=\exp(\eta_{ik})/\sum_{k'=1}^3 \exp(\eta_{ik'})$ for $k=1,2,3$, where $\eta_{i1}= 0.8X_i + 0.5Z_{i1} - 0.3Z_{i2}$, $\eta_{i2}= -0.4X_i + 0.2Z_{i1} + 0.4Z_{i2}$ and $\eta_{i3}= 0$.

The baseline function $\mu$ is specified as $\mu(x,z)= g_{\mathrm{int}}(z) + g_{\mathrm{slope}}(z)\,f(x)$, where the function $f$ is constructed to satisfy $f(0)=0$.
As a result, the baseline level at the cutoff is given by
$\mu(0,z)=g_{\mathrm{int}}(z)$.
The component functions are defined as  
\begin{align*}
&g_{\mathrm{int}}(z) = \alpha_{\mu}
\Big[ 1+\Phi\left(\frac{z_1 + 1}{2}\right) + 0.1\sin(\pi z_1)+\frac{1}{\pi} \tan^{-1}(z_2-1) + 0.5\delta_{1}(z_5)+\delta_{3}(z_5) 
\Big]\\
&g_{\mathrm{slope}}(z) = 2 + \frac{\exp(z_1)}{1+\exp(z_1)} + I(z_3<0)\sqrt{|z_3|} + \max\{0, z_4\} + \delta_{3}(z_5),
\end{align*}
and $f(x)=x+ \sin(2\pi x)$, where $\Phi$ is the cumulative probability function of the standard normal distribution and $\delta_a(z_5)=I(z_5=a)$ with $I$ being the indicator function. 
The parameter $\alpha_{\mu}$ directly controls the magnitude of baseline variation at the cutoff through $\mathrm{Var}(\mu(0,Z)\mid X=0)$.
The treatment effect function $\tau$ is given by
$$
\tau(z)=\alpha_{\tau}\Big\{1+0.5\cos(2\pi z_1)+0.6z_2z_3+0.4z_4-0.5I(z_5=2)\Big\}.
$$
The scaling parameter $\alpha_{\tau}$ is chosen so that
$\mathrm{Var}\bigl(\tau(Z)\mid X=0\bigr) = 0.5$.
We consider two cases that differ in the magnitude of baseline variation: (a) $\mathrm{Var}(\mu(0,Z)\mid X=0)=1$ (small baseline variability) and (b) $\mathrm{Var}(\mu(0,Z)\mid X=0)=15$ (large baseline variability). 
For each case, we vary the noise variance as $\sigma^2\in\{0.25,0.5,1\}$.
Under small baseline variability, variation in treatment effects accounts for a substantial fraction of the total outcome variability, making CATE estimation straightforward.
In contrast, in the Hard case, baseline variation dominates outcome variability, leading to a low signal-to-noise ratio for treatment effect estimation and making CATE estimation difficult.

\paragraph{{\bf (Scenario 2)}} \
We adopt a data-generating process originally introduced in
\citet{alcantara2025learning}.
We set $n=600$ and generate $4$-dimensional covariate vector $Z_i = (Z_{i1}, \ldots, Z_{i4})^\top$ from a multivariate normal distribution, $Z_i \sim \mathcal N_4(0, \Sigma_Z)$, where $\Sigma_Z$ is a Toeplitz covariance matrix with entries ranging from 0 to 2.
The running variable $X_i$ is generated from a normal distribution.
$X_i\mid Z_i \sim \mathcal{N}(\gamma_0 + \gamma^{\top}Z_i, \nu)$, where $\gamma$ is a coefficient vector with identical components. By shifting the mean toward the treatment region through the additive constant $\gamma_0 = 1$, the resulting distribution of the running variable is asymmetric around the cutoff at 0.
The key parameter is $\rho$, which controls the strength of the correlation between the running variable and the covariates. 
Given a specified value of $\rho$, the parameters $\gamma$ and $\nu$ are determined so as to satisfy $\mathrm{Var}(X_i) = 1$ and $\mathrm{Cor}(X_i, \gamma^{\top} Z_i) = \rho$.
The baseline function $\mu$ is constructed as $\mu(x,z) = \beta_{\mu} \mu^{\star}(x,z)$, where the scaling constant $\beta_{\mu}$ is chosen so that $\mathrm{Var}(\mu(0,Z)\mid X=0) = 1$. The unscaled function $\mu^{\star}(x,z)$ is given by  
$$
\mu^{\star}(x,z) = (x+1)^3 + (z^{\star} + 2)^2\Big(\text{sign}(x+1)\sqrt{|x+1|}\Big),
$$
where $z^{\star} = \sum_{j = 1}^4 z_{j}/\sqrt{4}$.
The treatment effect function $\tau$ is defined as $\tau(z) = \alpha_{\tau} + \beta_{\tau} \tau^{\star}(z)$, where the shifting parameter $\alpha_{\tau}$ and scaling parameter $\beta_{\tau}$ are 
fixed to satisfy $\sqrt{\mathrm{Var}(\tau(Z)\mid X=0)}= 1$ and $ \underset{Z\mid X=0}{\operatorname{min}}  \ \tau(Z)= 0$. 

The unscaled function $\tau^{\star}(z)$ is given by
$$
\tau^{\star}(z) = \frac{1}{2}\Phi(2z_1+3) + \phi(z_1),
$$
where $\Phi$ and $\phi$ are the cumulative probability function and the probability density function of the standard normal distribution.
We consider three choices of $\rho$: (a) $\rho=0$ (no correlation);
(b) $\rho = 0.25$ (low correlation); and (c) $\rho = 0.5$ (high correlation).
In all settings, we vary the noise variance as $\sigma^2\in\{0.5,1\}$.

\subsection{Comparative Methods and Evaluation Criterion}
For the simulated dataset, we applied the following methods: 
\begin{itemize}
\item[-] {\bf Direct-BART} (proposed): It applies the BART (with 20 trees) directly to CATE through a local loss function around the cutoff. We employ a local polynomial of order $q=2$ for Scenario 1 and $q=1$ for Scenario 2, guided by exploratory inspection of the
relationship between the outcome and the running variable. 
For the bandwidth adaptation, we construct a candidate set of bandwidths by taking six equally spaced values over the interval from 0 to twice the bandwidth selected by the local polynomial estimator. To compute the Hyv\"{a}rinen score (\ref{H-score}), we use $500$ posterior samples after discarding the first $500$ samples as burn-in. 

\item[-] {\bf BARDDT}: 
This method is a BART-based approach proposed by \cite{alcantara2025learning}, in which the data are recursively partitioned using the running variable and covariates. Within each leaf, a piecewise linear function that is allowed to jump at the cutoff is fitted.
This method is implemented by using R package "stochtree" with 20 and 50 trees. 

\item[-] {\bf S-BART}: This method corresponds to the S-learner implementation by applying BART (with 20 trees) for modeling conditional expectation of the outcome, in which the treatment indicator $W_i$ is included as an additional covariate. This method is implemented by using R package "stochtree". 

\item[-] {\bf LP}: Local polynomial estimator implemented by using R package "rdrobust" without covariates. Since it only provides average treatment effect, the resulting CATE estimate is constant. 
\end{itemize}

For Direct-BART, we generated 4500 posterior samples after discarding the first 500 samples as burn-in in both Scenarios. 
For the other BART-based methods, we generated 10000 posterior samples after discarding the first 1000 samples as burn-in in Scenario 1, following the MCMC settings in \citet{StarlingMurrayCarvalhoBukowskiScott2020BARTSmoothing}. 
In Scenario 2, we first ran 30 grow-from-root (GFR) iterations to obtain an initial set of trees, 
and then generated 1000 posterior samples, following the MCMC settings in \citet{alcantara2025learning}.
 Note that the local polynomial estimator without covariates provides an ATE estimator and serves as a baseline to illustrate the gains from explicitly modeling conditional treatment effect heterogeneity.
We do not include the T-learner implementation based on BART (T-BART) in our comparisons. T-BART fits two separate BART models to the treated and control groups, respectively, and estimates the treatment effect as the difference between the predicted outcomes at the cutoff.
Because prediction at the cutoff necessarily involves extrapolation, and tree-based methods generally perform poorly in extrapolation tasks, we exclude T-BART from the comparison.

We evaluate both in-sample and out-of-sample performance of each method using the root mean squared error (RMSE) and the coverage probability of the $95\%$ credible intervals.
For the out-of-sample evaluation, $200$ target units are newly generated from the conditional distribution $Z \mid X = 0$. 
In contrast, the in-sample evaluation is conducted using observed units in the original sample.
Following \citet{alcantara2025learning}, we consider the CATE function defined as $\tau(Z_i) \equiv \mathbb{E}[Y_i(1) - Y_i(0) \mid Z_i]$.
Since observations far from the cutoff $X = c$ contribute little to identification of the treatment effect, we restrict attention to units within a neighborhood of the cutoff. 
Specifically, we evaluate $\tau(Z_i)$ for $i \in \{ i : |X_i - c| \le 0.1 \mathrm{sd}(X)\}$, where $\mathrm{sd}(X) $ is the standard deviation of $X_i$. 
All results are based on $15$ replications for each setting under the two scenarios.

\subsection{Results}
\begin{figure}
    \centering
\includegraphics[
  width=\linewidth,
  keepaspectratio]{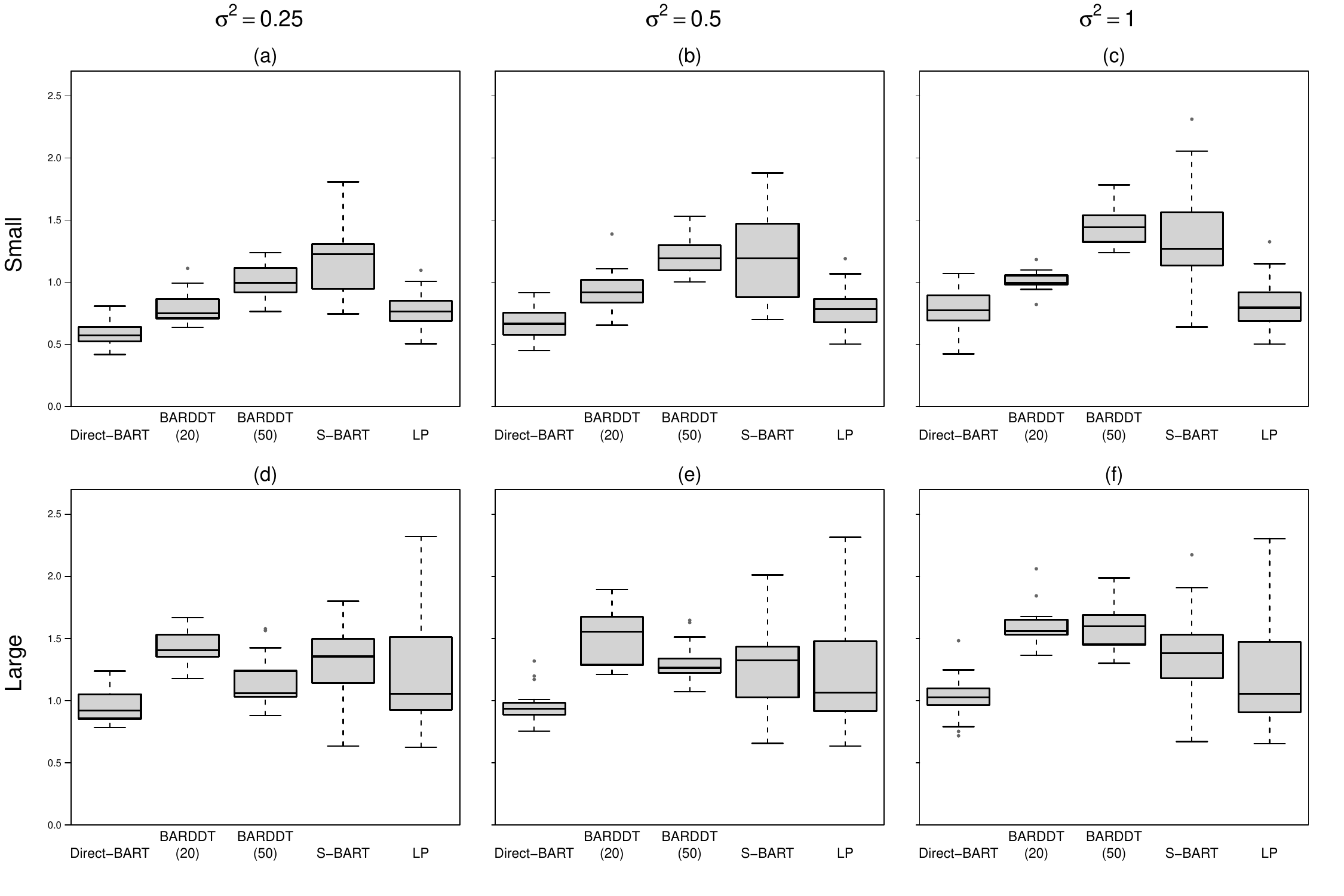}
    \caption{Boxplots of RMSE for in-sample units based on 15 replications for four methods under Scenario 1. The upper row corresponds to the small baseline variability case, while the lower row corresponds to the large baseline variability case.
    The left, middle, and right columns report results for noise variances $\sigma^2 = 0.25$, $0.5$, and $1$, respectively.
    }
    \label{fig:DGP1}
\end{figure}

\begin{figure}
\centering
\includegraphics[width=\linewidth]{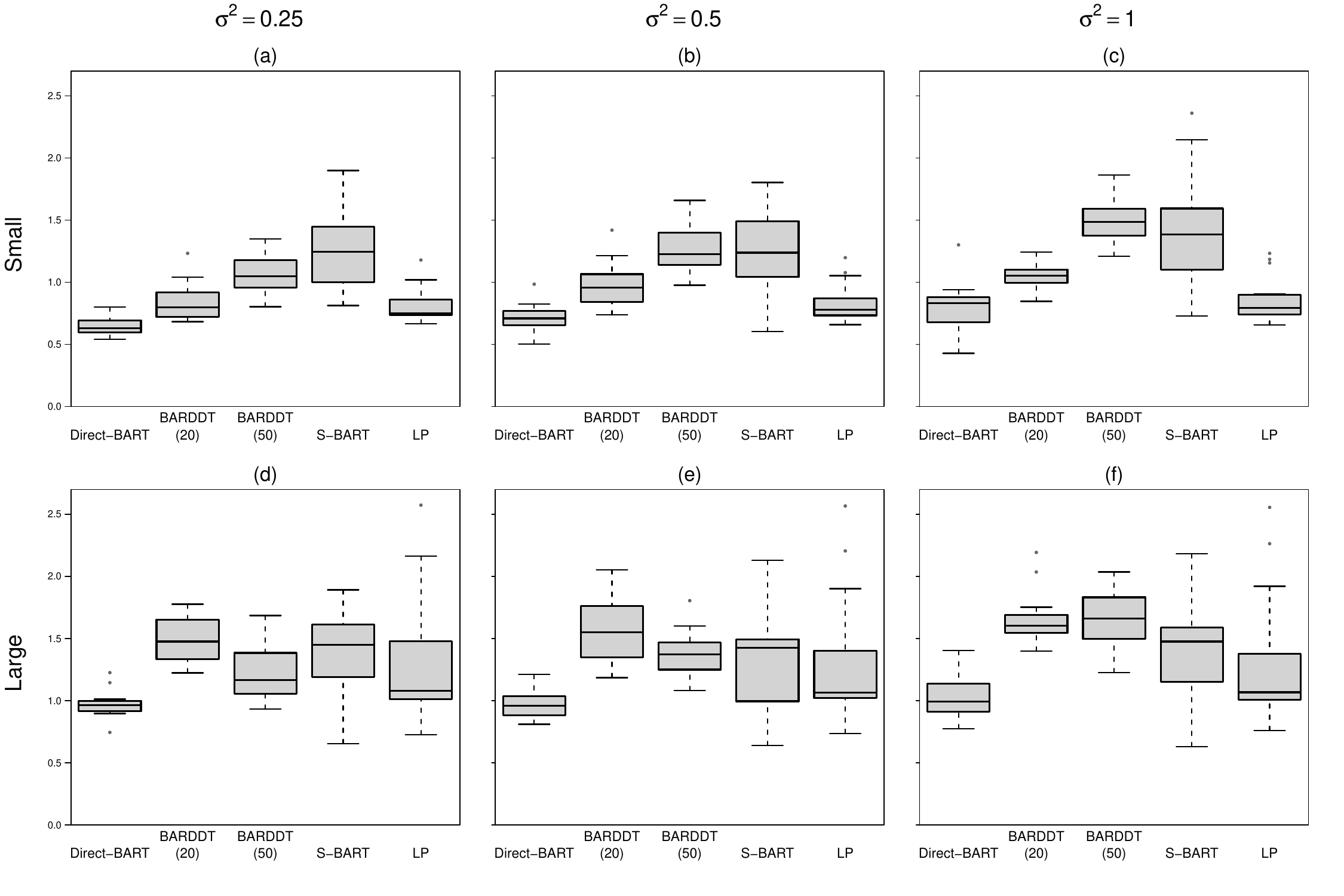}
\caption{Boxplots of RMSE for out-of-sample units based on 15 replications for four methods under Scenario~1. The upper row corresponds to the small baseline variability case, while the lower row corresponds to the large baseline variability case.
The left, middle, and right columns report results for noise variances $\sigma^2 = 0.25$, $0.5$, and $1$, respectively.
}
    \label{fig:testDGP1}
\end{figure}

We first show the results in Scenario~1.
Figure~\ref{fig:DGP1} displays boxplots of the RMSE for the four methods evaluated on the in-sample units across three noise variance levels, separately for the small and large baseline variability cases. Figure~\ref{fig:testDGP1} shows the corresponding results evaluated on the out-of-sample units.
In the small variability case, the proposed method attains lower RMSEs than competing methods in all settings except case (c).
In the high-noise setting (c), the proposed method exhibits performance comparable to that of the local polynomial estimator and S-BART.
In the large variability case, the proposed method generally achieves lower RMSEs than competing methods, regardless of the noise variance.
Overall, these results indicate that the proposed method provides accurate estimation in terms of RMSE under Scenario 1. 
Table~\ref{tab:summary_dgp1} reports average RMSE of point estimates and average coverage probability of $95\%$ credible (confidence) intervals for four methods across three noise variance levels, reported separately for the small and large baseline variability cases.
In terms of interval estimation, Direct-BART improves coverage over other BART-based alternatives, although some under-coverage remains, especially in the large-variability settings. Overall, these results suggest that Direct-BART provides more reliable interval estimation than the competing BART-based methods under Scenario 1.

\sisetup{detect-all}
\begin{table}[htbp]
\centering
\caption{RMSE of point estimates and coverage probability of $95\%$ credible (confidence) intervals, averaged over 15 Monte Carlo replications, under Scenario~1.}
\label{tab:summary_dgp1}
\begin{threeparttable}
\setlength{\tabcolsep}{6pt}
\renewcommand{\arraystretch}{1.15}

\begin{tabular}{l l *{6}{r}}
\toprule
& Case
& \multicolumn{3}{c}{small variability}
& \multicolumn{3}{c}{large variability} \\
\cmidrule(lr){3-5}\cmidrule(lr){6-8}
& $\sigma^2$
& {0.25} & {0.5} & {1.0}
& {0.25} & {0.5} & {1.0} \\
\midrule

% ================= In-sample RMSE =================
\multirow{5}{*}{{\textbf{RMSE (in-sample)}}}
& Direct-BART       & 0.59 & 0.66 & 0.78 & 0.95 & 0.97 & 1.03 \\
& BARDDT (20)       & 0.80 & 0.94 & 1.01 & 1.43 & 1.51 & 1.61 \\
& BARDDT (50)       & 1.01 & 1.22 & 1.47 & 1.15 & 1.31 & 1.60 \\
& S-BART            & 1.18 & 1.21 & 1.37 & 1.32 & 1.26 & 1.37 \\
& LP                & 0.78 & 0.79 & 0.83 & 1.28 & 1.27 & 1.27 \\
\midrule

% ================= Out-of-sample RMSE =================
\multirow{5}{*}{{\textbf{RMSE (out-of-sample)}}}
& Direct-BART       & 0.64 & 0.72 & 0.80 & 0.97 & 0.98 & 1.03 \\
& BARDDT (20)       & 0.85 & 0.97 & 1.04 & 1.48 & 1.56 & 1.66 \\
& BARDDT (50)       & 1.06 & 1.26 & 1.50 & 1.22 & 1.38 & 1.66 \\
& S-BART            & 1.24 & 1.27 & 1.40 & 1.39 & 1.31 & 1.41 \\
& LP                & 0.82 & 0.83 & 0.86 & 1.31 & 1.30 & 1.30 \\
\midrule

% ================= In-sample CI =================
\multirow{5}{*}{\shortstack{\textbf{Coverage probability (\%)}\\\textbf{(in-sample)}}}
& Direct-BART       & 88.9 & 91.1 & 93.9 & 74.9 & 83.2 & 90.2 \\
& BARDDT (20)       & 50.4 & 42.5 & 42.1 & 34.7 & 36.5 & 33.4 \\
& BARDDT (50)       & 41.6 & 35.6 & 28.6 & 48.9 & 44.1 & 38.8 \\
& S-BART            & 41.5 & 51.5 & 51.7 & 36.5 & 43.9 & 49.7 \\
& LP                & 75.8 & 77.0 & 79.2 & 93.4 & 93.1 & 93.0 \\
\midrule

% ================= Out-of-sample CI =================
\multirow{5}{*}{\shortstack{\textbf{Coverage probability (\%)}\\\textbf{(out-of-sample)}}}
& Direct-BART       & 88.3 & 91.6 & 94.2 & 75.6 & 83.7 & 91.7 \\
& BARDDT (20)       & 48.9 & 43.1 & 40.6 & 34.8 & 34.9 & 34.4 \\
& BARDDT (50)       & 41.4 & 34.8 & 26.8 & 47.0 & 43.0 & 37.5 \\
& S-BART            & 39.1 & 50.3 & 51.6 & 34.6 & 43.4 & 47.9 \\
& LP                & 74.7 & 76.6 & 79.0 & 92.9 & 92.9 & 92.8 \\
\bottomrule
\end{tabular}
\end{threeparttable}
\end{table}

\begin{figure}[htbp]
\centering
\includegraphics[width=\linewidth]{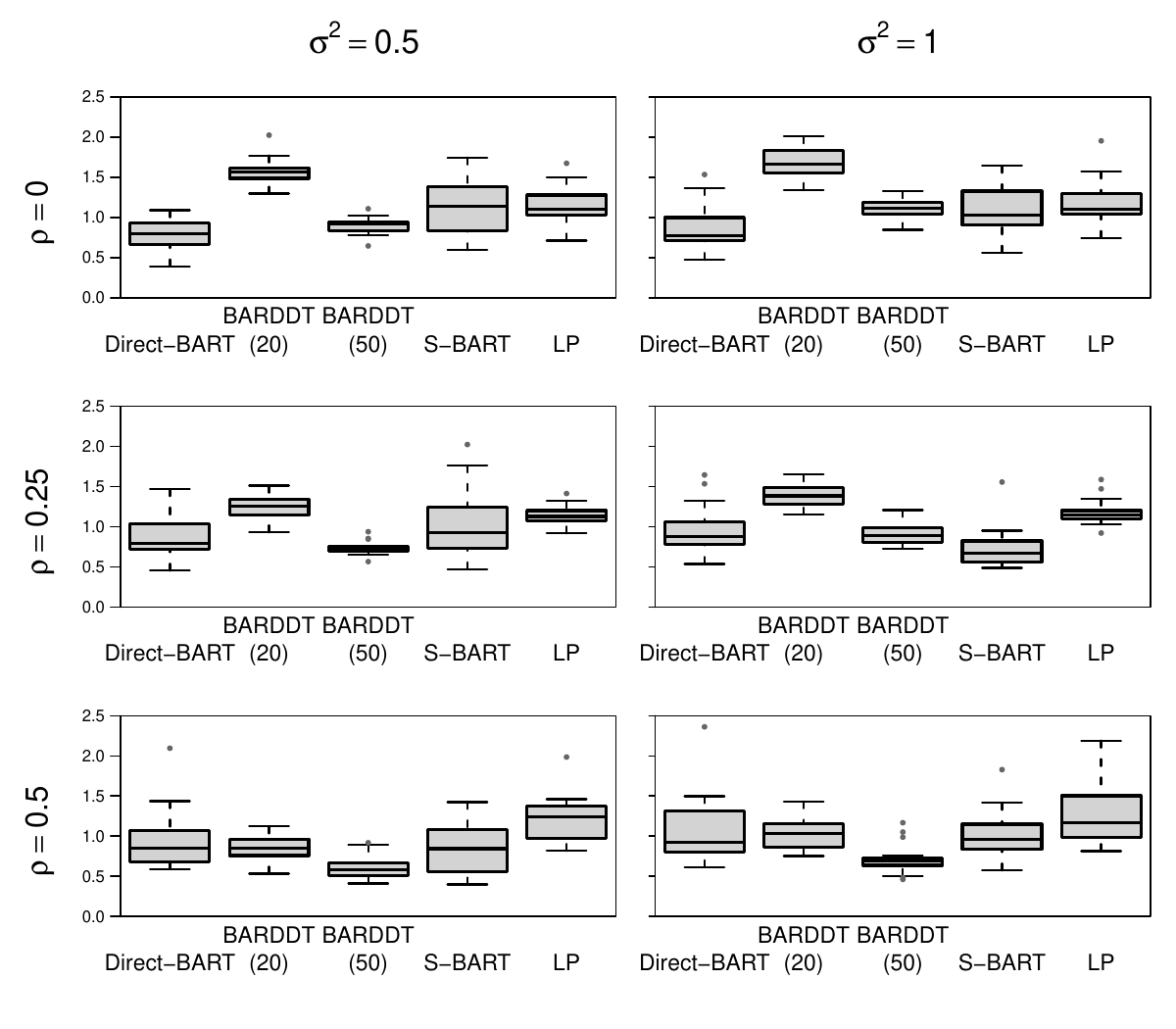}
\caption{Boxplots of RMSE for in-sample units based on 15 replications for four methods under Scenario~2.
The top, middle, and bottom panels correspond to the settings
$\rho = 0$, $\rho = 0.25$, and $\rho = 0.5$, respectively.
For each row, results are reported separately for
noise variances $\sigma^2 = 0.5$ and $\sigma^2 = 1$.
}
\label{fig:DGP2}
\end{figure}

\begin{figure}[htbp]
\centering
\includegraphics[width=\linewidth]{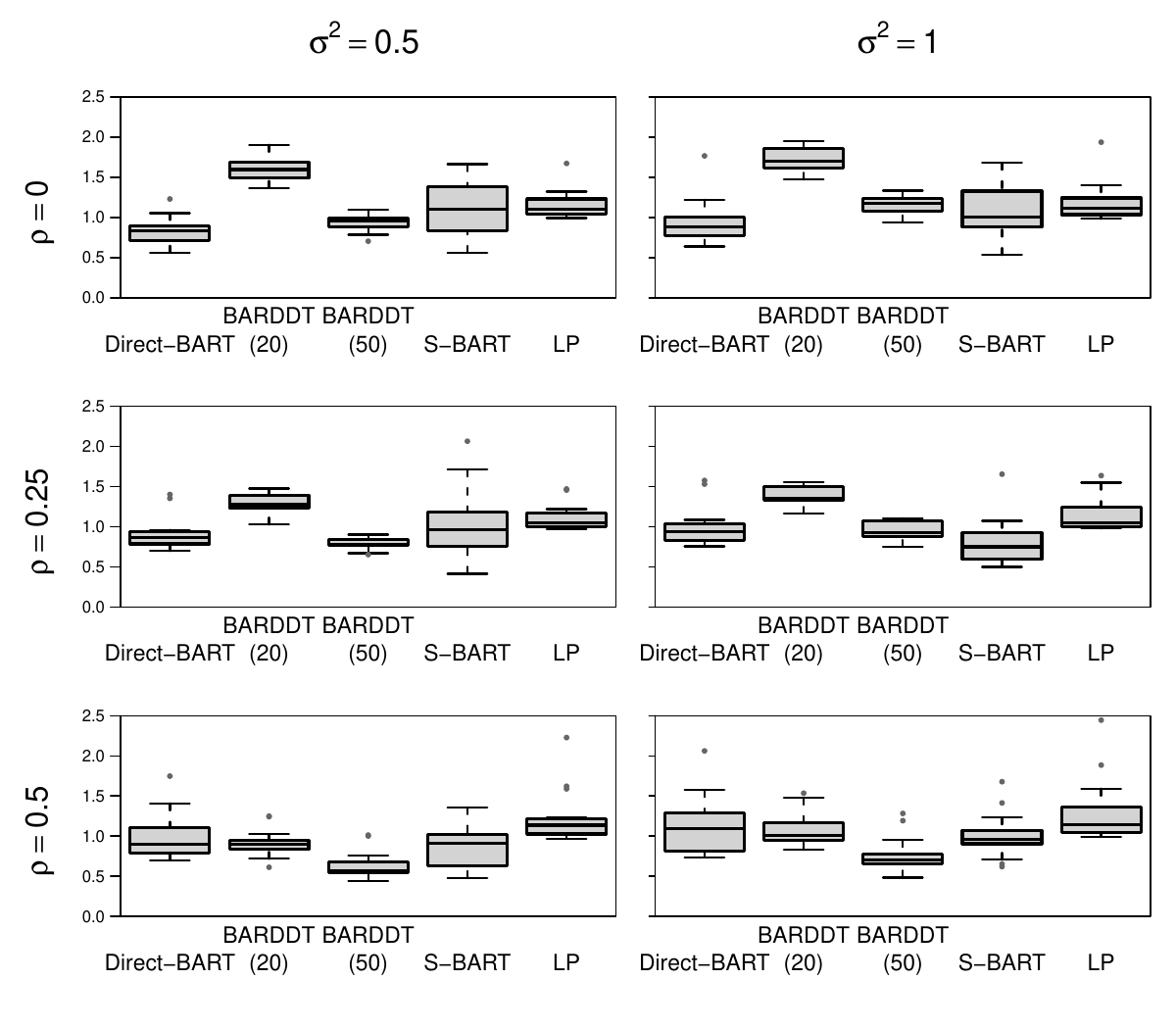}
\caption{Boxplots of RMSE for out-of-sample units based on 15 replications for four methods under Scenario~2.
The top, middle, and bottom panels correspond to the settings
$\rho = 0$, $\rho = 0.25$, and $\rho = 0.5$, respectively.
For each row, results are reported separately for
noise variances $\sigma^2 = 0.5$ and $\sigma^2 = 1$.
}
\label{fig:testDGP2}
\end{figure}

We next present the results under Scenario 2. 
Figure~\ref{fig:DGP2} displays boxplots of the RMSE for the four methods evaluated on the in-sample units across three correlation levels, reported separately for two noise variance levels.
Figure~\ref{fig:testDGP2} shows the corresponding results evaluated on the out-of-sample units. 
In the setting $\rho = 0$, Direct-BART achieves lower RMSEs than the other methods in terms of both the in-sample and out-of-sample evaluations. 
As $\rho$ increases, the advantage of Direct-BART in RMSE becomes less pronounced; nevertheless, it remains competitive across all settings. These results indicate that Direct-BART maintains robust point estimation performance under Scenario~2.
Table~\ref{tab:summary_dgp2} reports average RMSE of point estimates and average coverage probability of $95\%$ credible (confidence) intervals across three correlation levels, reported separately for two noise variance levels. 
Regarding the interval estimation, Direct-BART maintains relatively stable coverage across all settings, although some under-coverage remains, particularly when $\rho$ is large. As a whole, these results suggest that Direct-BART provides more stable interval estimation than the competing BART-based methods under Scenario 2. 
Taken together with evaluation in terms of point estimation and interval estimation, Direct-BART provides a favorable balance between point estimation accuracy and interval estimation stability under Scenario 2.


\sisetup{detect-all}
\begin{table}[htbp]
\centering
\caption{RMSE of point estimates and coverage probability of $95\%$ credible (confidence) intervals, averaged over 15 Monte Carlo replications, under Scenario 2.}
\label{tab:summary_dgp2}
\begin{threeparttable}
\setlength{\tabcolsep}{6pt}
\renewcommand{\arraystretch}{1.15}

\begin{tabular}{l l *{6}{r}}
\toprule
& $\rho$
& \multicolumn{2}{c}{$0$}
& \multicolumn{2}{c}{$0.25$}
& \multicolumn{2}{c}{$0.5$} \\
\cmidrule(lr){3-4}\cmidrule(lr){5-6}\cmidrule(lr){7-8}
& $\sigma^2$
& {0.5} & {1}
& {0.5} & {1}
& {0.5} & {1} \\
\midrule

% ================= In-sample RMSE =================
\multirow{5}{*}{{\textbf{RMSE (in-sample)}}}
& Direct-BART       & 0.79 & 0.89 & 0.86 & 0.97 & 0.96 & 1.11 \\
& BARDDT (20)       & 1.57 & 1.68 & 1.25 & 1.38 & 0.86 & 1.04 \\
& BARDDT (50)       & 0.90 & 1.11 & 0.73 & 0.91 & 0.61 & 0.72 \\
& S-BART            & 1.12 & 1.09 & 1.02 & 0.73 & 0.86 & 1.03 \\
& LP                & 1.15 & 1.18 & 1.14 & 1.18 & 1.21 & 1.28 \\
\midrule

% ================= Out-of-sample RMSE =================
\multirow{5}{*}{{\textbf{RMSE (out-of-sample)}}}
& Direct-BART       & 0.84 & 0.93 & 0.91 & 1.00 & 1.00 & 1.13 \\
& BARDDT (20)       & 1.61 & 1.72 & 1.29 & 1.40 & 0.91 & 1.09 \\
& BARDDT (50)       & 0.93 & 1.15 & 0.79 & 0.96 & 0.64 & 0.76 \\
& S-BART            & 1.11 & 1.09 & 1.03 & 0.80 & 0.86 & 1.00 \\
& LP                & 1.16 & 1.18 & 1.12 & 1.15 & 1.23 & 1.29 \\
\midrule

% ================= In-sample CI =================
\multirow{5}{*}{\shortstack{\textbf{Coverage probability (\%)}\\\textbf{(in-sample)}}}
& Direct-BART       & 91.0 & 92.8 & 90.1 & 93.1 & 85.2 & 84.0 \\
& BARDDT (20)       & 15.3 & 14.7 & 23.0 & 19.4 & 44.9 & 35.4 \\
& BARDDT (50)       & 34.9 & 30.8 & 49.9 & 45.2 & 72.6 & 71.1 \\
& S-BART            & 66.1 & 73.0 & 79.1 & 96.3 & 94.9 & 96.6 \\
& LP                & 65.6 & 70.1 & 72.7 & 76.5 & 66.1 & 66.7 \\
\midrule

% ================= Out-of-sample CI =================
\multirow{5}{*}{\shortstack{\textbf{Coverage probability (\%)}\\\textbf{(out-of-sample)}}}
& Direct-BART       & 91.8 & 94.1 & 91.5 & 92.7 & 86.1 & 87.5 \\
& BARDDT (20)       & 14.9 & 13.1 & 23.7 & 22.5 & 41.0 & 34.6 \\
& BARDDT (50)       & 35.5 & 31.7 & 47.3 & 42.8 & 71.2 & 67.0 \\
& S-BART            & 67.1 & 74.0 & 79.6 & 94.8 & 96.5 & 94.4 \\
& LP                & 67.6 & 73.4 & 75.0 & 78.0 & 65.6 & 65.9 \\
\bottomrule
\end{tabular}
\end{threeparttable}
\end{table}

%----------------------------------------%
%         Concluding Remarks             %
%----------------------------------------%
\section{Concluding Remarks}
\label{sec:Concluding Remarks}
In this paper, we propose a Bayesian nonparametric approach to estimating CATE in RDD based on BART. 
Through extensive simulation studies under both newly designed and previously studied DGPs, we showed that the proposed method consistently achieves favorable RMSE performance relative to existing methods, as it can capture complicated structures of heterogeneous treatment effects.

Although we employed the standard BART prior in this paper, a variety of extensions and computational improvements have been developed in the literature.
For example, scalable variants and regularized formulations of BART have been proposed to handle high-dimensional covariates \citep{linero2018bayesian} and large datasets \citep{he2019xbart}.
Incorporating such advanced BART variants into the proposed framework would be a promising direction for future work, particularly in settings with high-dimensional covariates or complex treatment effect heterogeneity.
Another important extension is to consider fuzzy RDD settings \citep[e.g.][]{ImbensLemieux2008RDDGuide}, where treatment assignment is not perfectly determined by the running variable.
Extending the proposed framework to accommodate fuzzy designs, for example through instrumental-variable formulations within the BART framework, would broaden the applicability of the method and is left for future research.

%--------------------------------------------%
%           Acknowledgement                  %
%--------------------------------------------%
\section*{Acknowledgement}
This work is partially supported by the Japan Society for the Promotion of Science (JSPS KAKENHI) grant numbers 24K21420 and 25H00546.

%  Reference 
\bibliographystyle{chicago}
\bibliography{refs-RDD}

\begin{thebibliography}{}

\bibitem[\protect\citeauthoryear{Alcantara, Hahn, Carvalho, and Lopes}{Alcantara et~al.}{2025}]{alcantara2025learning}
Alcantara, R., P.~R. Hahn, C.~Carvalho, and H.~Lopes (2025).
\newblock {L}earning {C}onditional {A}verage {T}reatment {E}ffects in {R}egression {D}iscontinuity {D}esigns using {B}ayesian {A}dditive {R}egression {T}rees.
\newblock {\em arXiv preprint arXiv:2503.00326\/}.

\bibitem[\protect\citeauthoryear{Becker, Egger, and Von~Ehrlich}{Becker et~al.}{2013}]{becker2013absorptive}
Becker, S.~O., P.~H. Egger, and M.~Von~Ehrlich (2013).
\newblock Absorptive capacity and the growth and investment effects of regional transfers: {A} regression discontinuity design with heterogeneous treatment effects.
\newblock {\em American Economic Journal: Economic Policy\/}~{\em 5\/}(4), 29--77.

\bibitem[\protect\citeauthoryear{Bissiri, Holmes, and Walker}{Bissiri et~al.}{2016}]{Bissiri2016GeneralFramework}
Bissiri, P.~G., C.~C. Holmes, and S.~G. Walker (2016).
\newblock A {G}eneral {F}ramework for {U}pdating {B}elief {D}istributions.
\newblock {\em Journal of the Royal Statistical Society: Series B (Statistical Methodology)\/}~{\em 78\/}(5), 1103--1130.

\bibitem[\protect\citeauthoryear{Calonico, Cattaneo, Farrell, Palomba, and Titiunik}{Calonico et~al.}{2025}]{calonico2025treatment}
Calonico, S., M.~D. Cattaneo, M.~H. Farrell, F.~Palomba, and R.~Titiunik (2025).
\newblock Treatment effect heterogeneity in regression discontinuity designs.
\newblock {\em arXiv preprint arXiv:2503.13696\/}.

\bibitem[\protect\citeauthoryear{Calonico, Cattaneo, and Titiunik}{Calonico et~al.}{2014}]{CalonicoCattaneoTitiunik2014RDDCI}
Calonico, S., M.~D. Cattaneo, and R.~Titiunik (2014).
\newblock Robust {N}onparametric {C}onfidence {I}ntervals for {R}egression-{D}iscontinuity {D}esigns.
\newblock {\em Econometrica\/}~{\em 82\/}(6), 2295--2326.

\bibitem[\protect\citeauthoryear{Card, Dobkin, and Maestas}{Card et~al.}{2008}]{card2008impact}
Card, D., C.~Dobkin, and N.~Maestas (2008).
\newblock The impact of nearly universal insurance coverage on health care utilization: evidence from {M}edicare.
\newblock {\em American Economic Review\/}~{\em 98\/}(5), 2242--2258.

\bibitem[\protect\citeauthoryear{Chib, Greenberg, and Simoni}{Chib et~al.}{2023}]{ChibGreenbergSimoni2023NonparametricRD}
Chib, S., E.~Greenberg, and A.~Simoni (2023).
\newblock Nonparametric {B}ayes {A}nalysis of the {S}harp and {F}uzzy {R}egression {D}iscontinuity {D}esigns.
\newblock {\em Econometric Theory\/}~{\em 39\/}(3), 481--533.

\bibitem[\protect\citeauthoryear{Chipman, George, and McCulloch}{Chipman et~al.}{2010}]{Chipman2010BART}
Chipman, H.~A., E.~I. George, and R.~E. McCulloch (2010).
\newblock {BART}: {B}ayesian {A}dditive {R}egression {T}rees.
\newblock {\em The Annals of Applied Statistics\/}~{\em 4\/}(1), 266--298.

\bibitem[\protect\citeauthoryear{Hahn, Todd, and Van~der Klaauw}{Hahn et~al.}{2001}]{hahn2001identification}
Hahn, J., P.~Todd, and W.~Van~der Klaauw (2001).
\newblock Identification and estimation of treatment effects with a regression-discontinuity design.
\newblock {\em Econometrica\/}~{\em 69\/}(1), 201--209.

\bibitem[\protect\citeauthoryear{Hahn, Murray, and Carvalho}{Hahn et~al.}{2020}]{hahn2020bayesian}
Hahn, P.~R., J.~S. Murray, and C.~M. Carvalho (2020).
\newblock Bayesian regression tree models for causal inference: Regularization, confounding, and heterogeneous effects (with discussion).
\newblock {\em Bayesian Analysis\/}~{\em 15\/}(3), 965--1056.

\bibitem[\protect\citeauthoryear{He, Yalov, and Hahn}{He et~al.}{2019}]{he2019xbart}
He, J., S.~Yalov, and P.~R. Hahn (2019).
\newblock {XBART}: {A}ccelerated {B}ayesian additive regression trees.
\newblock In {\em The 22nd International Conference on Artificial Intelligence and Statistics}, pp.\  1130--1138. PMLR.

\bibitem[\protect\citeauthoryear{Hyv{\"a}rinen}{Hyv{\"a}rinen}{2005}]{HyvarinenDayan2005ScoreMatching}
Hyv{\"a}rinen, A. (2005).
\newblock Estimation of {N}on-{N}ormalized {S}tatistical {M}odels by {S}core {M}atching.
\newblock {\em Journal of Machine Learning Research\/}~{\em 6\/}(4), 695--709.

\bibitem[\protect\citeauthoryear{Imbens and Kalyanaraman}{Imbens and Kalyanaraman}{2012}]{ImbensKalyanaraman2012IK}
Imbens, G. and K.~Kalyanaraman (2012).
\newblock Optimal {B}andwidth {C}hoice for the {R}egression {D}iscontinuity {E}stimator.
\newblock {\em The Review of Economic Studies\/}~{\em 79\/}(3), 933--959.

\bibitem[\protect\citeauthoryear{Imbens and Lemieux}{Imbens and Lemieux}{2008}]{ImbensLemieux2008RDDGuide}
Imbens, G.~W. and T.~Lemieux (2008).
\newblock {R}egression {D}iscontinuity {D}esigns: {A} {G}uide to {P}ractice.
\newblock {\em Journal of Econometrics\/}~{\em 142\/}(2), 615--635.

\bibitem[\protect\citeauthoryear{Lee and Lemieux}{Lee and Lemieux}{2010}]{LeeLemieux2010RDDSurvey}
Lee, D.~S. and T.~Lemieux (2010).
\newblock Regression {D}iscontinuity {D}esigns in {E}conomics.
\newblock {\em Journal of Economic Literature\/}~{\em 48\/}(2), 281--355.

\bibitem[\protect\citeauthoryear{Linero}{Linero}{2018}]{linero2018bayesian}
Linero, A.~R. (2018).
\newblock Bayesian regression trees for high-dimensional prediction and variable selection.
\newblock {\em Journal of the American Statistical Association\/}~{\em 113\/}(522), 626--636.

\bibitem[\protect\citeauthoryear{Reguly}{Reguly}{2021}]{reguly2021discovering}
Reguly, {\'A}. (2021).
\newblock Discovering {H}eterogeneous {T}reatment {E}ffects in {R}egression {D}iscontinuity {D}esigns.
\newblock {\em arXiv preprint arXiv:2106.11640\/}.

\bibitem[\protect\citeauthoryear{Sawada, Ishihara, Kurisu, and Matsuda}{Sawada et~al.}{2024}]{sawada2024local}
Sawada, M., T.~Ishihara, D.~Kurisu, and Y.~Matsuda (2024).
\newblock Local-polynomial estimation for multivariate regression discontinuity designs.
\newblock {\em arXiv preprint arXiv:2402.08941\/}.

\bibitem[\protect\citeauthoryear{Shao, Jacob, Ding, and Tarokh}{Shao et~al.}{2019}]{ShaoJacobDingTarokh2019Hyvarinen}
Shao, S., P.~E. Jacob, J.~Ding, and V.~Tarokh (2019).
\newblock Bayesian {M}odel {C}omparison with the {H}yv{\"a}rinen {S}core: {C}omputation and {C}onsistency.
\newblock {\em Journal of the American Statistical Association\/}~{\em 114\/}(528), 1826--1837.

\bibitem[\protect\citeauthoryear{Starling, Murray, Carvalho, Bukowski, and Scott}{Starling et~al.}{2020}]{StarlingMurrayCarvalhoBukowskiScott2020BARTSmoothing}
Starling, J.~E., J.~S. Murray, C.~M. Carvalho, R.~K. Bukowski, and J.~G. Scott (2020).
\newblock {BART} with {T}argeted {S}moothing: {A}n {A}nalysis of {P}atient-{S}pecific {S}tillbirth {R}isk.
\newblock {\em The Annals of Applied Statistics\/}.

\bibitem[\protect\citeauthoryear{Sugasawa, Ishihara, and Kurisu}{Sugasawa et~al.}{2023}]{sugasawa2023hierarchical}
Sugasawa, S., T.~Ishihara, and D.~Kurisu (2023).
\newblock Hierarchical regression discontinuity design: Pursuing subgroup treatment effects.
\newblock {\em arXiv preprint arXiv:2309.01404\/}.

\bibitem[\protect\citeauthoryear{Tao, Wang, and Ruppert}{Tao et~al.}{2025}]{tao2025bayesian}
Tao, K., Y.~S. Wang, and D.~Ruppert (2025).
\newblock Bayesian analysis of regression discontinuity designs with heterogeneous treatment effects.
\newblock {\em arXiv preprint arXiv:2504.10652\/}.

\bibitem[\protect\citeauthoryear{Thistlethwaite and Campbell}{Thistlethwaite and Campbell}{1960}]{ThistlethwaiteCampbell1960RDD}
Thistlethwaite, D.~L. and D.~T. Campbell (1960).
\newblock Regression-{D}iscontinuity {A}nalysis: {A}n {A}lternative to the {E}x {P}ost {F}acto {E}xperiment.
\newblock {\em Journal of Educational Psychology\/}~{\em 51\/}(6), 309--317.

\bibitem[\protect\citeauthoryear{Yonekura and Sugasawa}{Yonekura and Sugasawa}{2023}]{YonekuraSugasawa2023GeneralBayesRobust}
Yonekura, S. and S.~Sugasawa (2023).
\newblock Adaptation of the {T}uning {P}arameter in {G}eneral {B}ayesian {I}nference with {R}obust {D}ivergence.
\newblock {\em Statistics and Computing\/}~{\em 33\/}(2), 39.

\end{thebibliography}

\end{document}